# Long-Lived Charge Separation Following Pump-Energy Dependent Ultrafast Charge Transfer in Graphene/WS$_2$ Heterostructures

Shuai Fu[1], Indy du Fossé[2], Xiaoyu Jia[1], Jingyin Xu[1,3], Xiaoqing Yu[1], Heng Zhang[1], Wenhao Zheng[1], Sven Krasel[1], Zongping Chen[4], Zhiming M. Wang[3], Klaas-Jan Tielrooij[5], Mischa Bonn[1], Arjan J. Houtepen[2], Hai I. Wang[1*]

[1] *Max Planck Institute for Polymer Research, Ackermannweg 10, D-55128 Mainz, Germany*

[2] *Optoelectronic Materials Section, Faculty of Applied Sciences, Delft University of Technology, Van der Maasweg 9, 2629 HZ Delft, The Netherlands*

[3] *Institute of Fundamental and Frontier Sciences, University of Electronic Science and Technology of China, Chengdu, 610054 P. R. China*

[4] *School of Materials Science and Engineering, Zhejiang University, 38 Zheda Road Hangzhou 310027, China*

[5] *Catalan Institute of Nanoscience and Nanotechnology (ICN2), BIST and CSIC, Campus UAB, Bellaterra, 08193 Barcelona, Spain*

[*]E-mail: wanghai@mpip-mainz.mpg.de

**Abstract**

Van der Waals heterostructures consisting of graphene and transition metal dichalcogenides (TMDCs) have recently shown great promise for high-performance optoelectronic applications. However, an in-depth understanding of the critical processes for device operation, namely interfacial charge transfer (CT) and recombination, has so far remained elusive. Here, we investigate these processes in graphene-$WS_2$ heterostructures, by complementarily probing the ultrafast terahertz photoconductivity in graphene and the transient absorption dynamics in $WS_2$ following photoexcitation. We find that CT across graphene-$WS_2$ interfaces occurs via photo-thermionic emission for sub-A-exciton excitation, and direct hole transfer from $WS_2$ to the valence band of graphene for above-A-exciton excitation. Remarkably, we observe that separated charges in the heterostructure following CT live extremely long: beyond 1 ns, in contrast to ~1 ps charge separation reported in previous studies. This leads to efficient photogating of graphene. These findings provide relevant insights to optimize further the performance of optoelectronic devices, in particular photodetection.

**Introduction**

Atomically thin layers, including graphene and monolayer TMDCs, represent a fascinating material class for electronic and optoelectronic applications. As a bond-free strategy, stacking these two-dimensional (2D) layers allows the production of artificial vdW heterostructures, which offer the prospect of discovering new synergetic electronic, optical, and magnetic phenomena (*1–4*). Thanks to the development of precise control over composition, layer numbers, stacking angles and sequences of the atomic layers, the last decade has witnessed the blossoming of novel concepts and high-performance devices based on vdW heterostructures (*5–7*). One notable example is that the integration of graphene and monolayer TMDCs enables sensitive photodetectors with high photoresponsivity ($R_{ph}$) up to $10^7$ A W$^{-1}$ at room temperature (*8*), comparable to state-of-the-art photodetectors based on graphene-quantum dots (g-QDs) hybrid system with $R_{ph}$ up to $10^8$ A W$^{-1}$ (*9*).

There has been a great and successful effort to increase the efficiency of graphene-TMDCs (g-TMDCs) photodetectors, but the understanding of the fundamental photophysics of these devices has remained elusive. Immediately following photoexcitation of bare graphene, the photogenerated, non-thermalized hot carriers can efficiently transfer their excess energy to other charge carriers within tens of femtoseconds (fs) via carrier-carrier scattering (*10–13*). This thermalization process leads to the formation of thermalized hot carriers with a well-defined electron temperature ($T_e$) following the Fermi-Dirac distribution (*14*, *15*). The thermalized hot carriers undergo a cooling process within a few ps via electron-phonon scattering (*16–20*). In g-TMDCs vdW heterostructures, TMDCs can serve as transport channels to harvest hot carriers (in principle for both non-thermalized and thermalized hot carriers) from graphene before the cooling process takes place. Indeed, pioneering device work (*21*, *22*) has shown that hot electron transfer (HET) contributes to the photocurrent generation at g-based vdW interfaces, which has been further confirmed by ultrafast spectroscopy studies (*23–25*). However, the mechanism of HET across vdW interfaces, and in particular whether HET occurs before or after thermalization, remains highly debated. For instance, in the device work by Massicotte, the photocurrent generation in a g-WSe$_2$-g vdW heterostructure following below-WSe$_2$-bandgap excitation is attributed to photo-thermionic emission, in which thermalized hot electrons with energy above the interfacial energy barrier can be injected into WSe$_2$ (with a quantum yield of ~1 %) (*21*). In sharp contrast, Chen et al. (*23*) recently proposed that HET at g-WS$_2$ interfaces competes with the thermalization in graphene and shows an extremely high quantum yield (~ 50 %). Furthermore, Yuan et al. (*24*) suggested an alternative model for HET at g-WS$_2$ interfaces, in which a direct excitation from graphene to WS$_2$ can take place via charge transfer (CT) states due to strong interfacial electronic coupling. It is apparent that further studies on HET across g-TMDCs interfaces are required to solve this debate. Furthermore, while the contribution of hot electrons to ET has been recognized, it remains unclear if the "cold" electrons, e.g. the valance band electrons in graphene, are involved in interfacial ET at g-TMDCs interfaces.

Along with the controversy on the ET mechanisms, the second puzzle regarding the (interfacial) charge carrier dynamics in g-TMDCs heterostructures lies in the lifetime of the separated charges. Recent spectroscopic studies (*23–24*) have reported a very fast (~ 1 ps) charge recombination (via back electron transfer process) at g-WS$_2$ interfaces. Such short charge separation lifetime seems in contradiction to the large $R_{ph}$ reported in the photodetectors based on g-TMDCs heterostructures. For other g-based hybrid photodetectors, e.g. g-PbS QDs photodetectors (*9*), an extremely long interfacial charge separation time (~ 20 ms) has been attributed to carrier trapping in the QDs. The long-lived interfacial charge separation establishes an electric field at vdW interfaces, leading to photoconductive gain and, thus, a high $R_{ph}$ in the photodetectors (the so-called "photogating effect", in analogy to field-effect gating). As the photoconductive gain (*G*) is linearly proportional to the charge separation lifetime ($\tau_{cs}$) (*26*), it remains unclear how efficient photodetectors can be realized in g-TMDCs heterostructures (*5, 27, 28*) with a reported $\tau_{cs}$ of ~ 1 ps.

In this report, aiming to provide a comprehensive understanding of the interfacial carrier dynamics in g-WS$_2$ vdW heterostructures, we measure complementarily the ultrafast photoconductivity dynamics in graphene by terahertz (THz) spectroscopy (**Figure 1A**) and the excited state dynamics in TMDCs by transient absorption (TA) spectroscopy (**Figure 1B**) following photoexcitation with a wide range of photon energies from 0.7 eV to 3.1 eV. The unique combination of THz and TA spectroscopies enables us to track the charge carriers dynamics in both donors and acceptors independently, and to identify and quantify the trapping process (if active) at the interfaces. For the forward ET process from graphene to $WS_2$, we unveil two distinctively different ET regimes when exciting the heterostructure below (0.7-2 eV) or above the A-exciton resonance (2-3.1 eV) of monolayer $WS_2$. Exciting below the $WS_2$ A-exciton transition, we find that a relatively inefficient (~1%) hot electron transfer (HET) via photo-thermionic emission governs the ET process, in which thermalized hot electrons in graphene are emitted over the energy barrier and transferred into $WS_2$ (**Figure 1C**). In the second regime where the photoexcitation is above the A-exciton resonance of $WS_2$, we report a more efficient (up to ~5%) direct hole transfer (DHT) mechanism, which involves the photogenerated holes in the valence band of $WS_2$ and electrons in the valence band of graphene (**Figure 1D**). Importantly, we find that the charge separation lifetime following ultrafast ET is remarkably long-lived, beyond 1 ns (limited by the scan range of the setup). This observation is in sharp contrast to the short-lived excited state dynamics (~1 ps) in $WS_2$ observed by TA, both in this study and in previous works (*23*, *24*). We rationalize the discrepancy between the carrier lifetime in the electron donor (graphene) and the acceptor ($WS_2$) by the presence of trap states at g-$WS_2$ interfaces. These trap states can effectively capture the electrons from the excited states of $WS_2$ within 1 ps (corresponding to the fast decay in TA dynamics) and further store them for over 1 ns (corresponding to the long-lived photoconductivity in graphene by terahertz study) before recombining with the holes in graphene. This leads to a long-lived photogating effect in graphene.

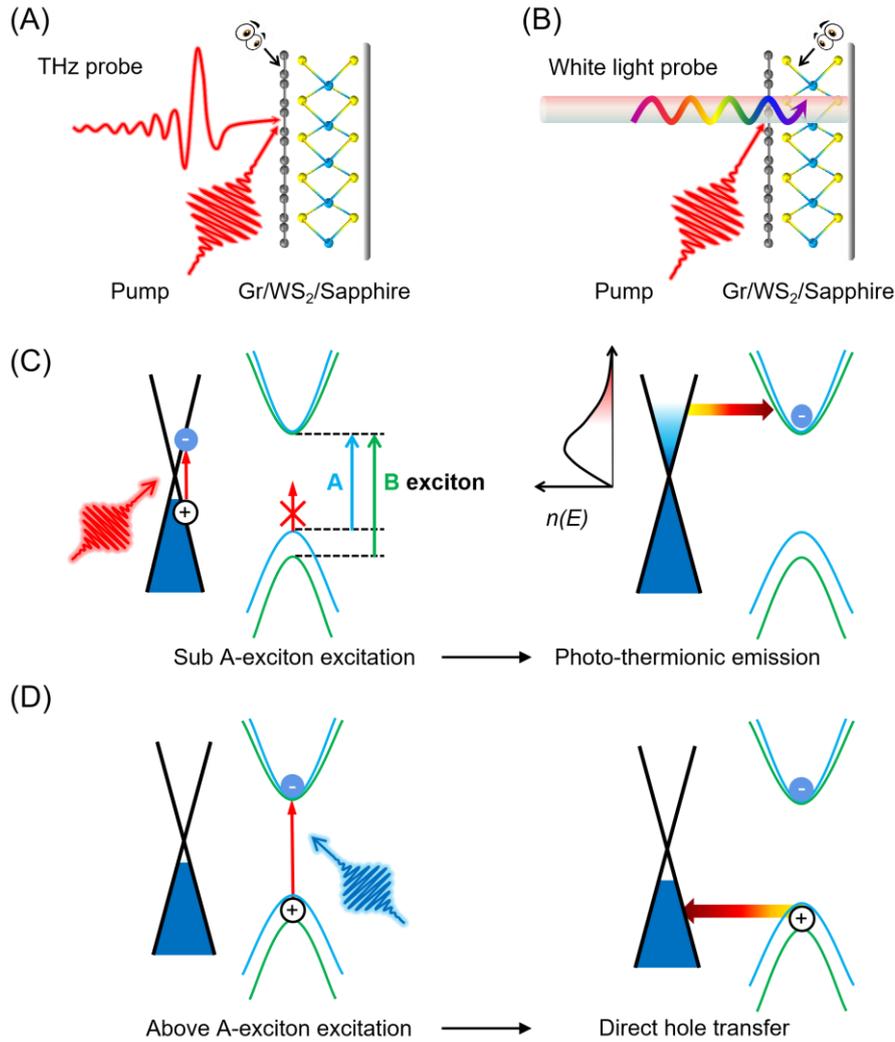

**Figure 1.** Investigation of the nonequilibrium hot carrier dynamics at g-WS$_2$ interfaces. (A) Schematic of employing ultrafast THz spectroscopy to measure the photoconductivity of graphene following photoexcitation. (B) Schematic of employing ultrafast TA spectroscopy to measure the excited state dynamics of WS$_2$ following photoexcitation; (C) Illustration of interfacial band alignment and photo-thermionic emission for thermalized hot electrons following sub-A-exciton excitation, in which thermalized hot electrons in graphene can be emitted over the energy barrier and transferred into the excited states of WS$_2$. A- and B-exciton transitions in WS$_2$, originating from the spin-split valence bands at the K point of the Brillouin zone, are marked. (D) Schematic of direct hole transfer (DHT) at g-WS$_2$ interfaces following above-A-exciton excitation, in which the photogenerated holes in the valance band of WS$_2$ recombine with valence band electrons in graphene.

## Results and discussion

The g-WS$_2$ heterostructure used in this study was obtained commercially (from SixCarbon Technology, Shenzhen). It is produced by transferring a chemical-vapor-deposited (CVD) graphene monolayer onto a CVD WS$_2$ monolayer, grown on a sapphire substrate (schematically depicted in **Figure S1(A)**). As both layers are multi-crystalline with a typical domain size of several micrometers, the dynamics that we are probing is an average response of the heterostructure with mixed twisting angles (given that the probe beam has a diameter of ~ 0.5 - 1 mm for THz measurements and ~ 0.1 mm for TA measurements). In an independent electrical four-point probe measurement of the resistivity of graphene (produced by the same method) at varied gating potentials, we find the graphene in the heterostructure is initially p-doped, in line with previous reports (*23*) for the same system. Prior to studying the ultrafast carrier dynamics, we have characterized the static optical and electronic properties of the heterostructure using UV-Vis absorption and Raman spectroscopy. We observe two exciton resonances at 2.0 eV and 2.4 eV in the absorption spectra (**Figure S1(B)**), corresponding to the A- and B-exciton transitions from the spin-split valence bands at the K point of the Brillouin zone in monolayer WS$_2$ (*29–31*), respectively (see also the simplified band structure of WS$_2$ in **Figure 1 (C-D)**). The featureless constant absorption (~2.3%) in the near-infrared range originates from the absorption of monolayer graphene (*32, 33*). Raman studies shown in **Figure S1(C)** and **Figure S1(D)** further confirm that both the graphene and WS$_2$ layer are monolayers in nature. Based on the G-band position in graphene, we can estimate the Fermi level ($E_F$) in graphene to be ~0.11 eV (equivalent to a free carrier density $N$ of $7.9 \times 10^{11}$ cm$^{-2}$, see **Section S1** in SI) below the Dirac point (given the p-doped nature of graphene from electrical measurements). We have further verified $E_F$ in graphene using THz-time domain spectroscopy (THz-TDS, see **Section S2** in SI). In short, in the THz-TDS measurements, we record the THz electric field transmitted through the WS$_2$/sapphire substrate configurations in the time domain, with and without graphene on the top, i.e., $E(t)$ (for graphene/WS$_2$/sapphire*)* and $E_0(t)$ (for WS$_2$/sapphire*)*, respectively. The obtained time-dependent THz fields are further converted into the frequency domain by Fourier transform as $E(\omega)$ and $E_0(\omega)$. The THz absorption*,* due to the presence of free charges in graphene, can be well described by the Drude model, which provides microscopic transport properties in graphene, including the charge carrier density (or Fermi level). Based on THz-TDS results, we estimate the $E_F$ in graphene to be 0.1 eV vs. the Dirac point (see **Figure S1(E-F)** and the associated discussions in **Section S2** in SI), in excellent agreement with our Raman measurements.

***Ultrafast interfacial charge transfer in g-WS$_2$ vdW heterostructures.*** We investigate the dynamics of photogenerated charge carriers in the g-WS$_2$ system by optical pump-terahertz probe (OPTP) spectroscopy. In a typical OPTP measurement, as shown in **Figure 1A**, an optical pump with ~50 fs duration selectively excites either only the graphene layer (sub-A-exciton transition in WS$_2$, $h\upsilon < 2$ eV) or both layers in the heterostructure at a fixed sample spot ($h\upsilon > 2$ eV). The pump-induced photoconductivity ($\Delta\sigma$) is probed by a THz pulse by monitoring the change in the transmitted electrical field ($\Delta E = E_{pump} - E_0$) following the optical pump as a function of pump-probe delay. The measurement is based on the principle that -$\Delta E$ is proportional to $\Delta\sigma$ (*34*).

As control measurements, we first investigate the carrier dynamics of the individual monolayer WS$_2$ and graphene upon 1.55 eV excitation. As the excitation energy is still below the A-exciton resonance of monolayer WS$_2$, we observe no photoconductivity (**Figure 2A**, blue line) for monolayer WS$_2$, as expected. On the other hand, the same excitation energy for graphene yields a transient reduction in the conductivity, i.e., negative photoconductivity (**Figure 2A**, grey circle points). The negative photoconductivity in doped graphene has been widely reported previously (*13*, *35–40*) and can be briefly understood as follows: following optical excitation and the rapid thermalization in the doped graphene, the increased carrier temperature leads to a reduced screening of the long-range Coulomb scattering, and thus reduced conductivity (*38*). Within a few ps, these thermalized hot carriers cool down to the initial thermal equilibrium via electron-phonon scattering.

For g-WS$_2$ heterostructures, one could, in the first instance, expect the photoconductivity upon 1.55 eV excitation to be a superposition of that of individual monolayer graphene and WS$_2$, i.e., exhibiting an overall negative photoconductivity with sub-10 ps lifetime. However, we observe distinctively different dynamics for the heterostructure upon 1.55 eV excitation (**Figure 2A**, red line): after the short-lived negative photoconductivity, the photoconductivity turns positive within 10 ps, followed by a remarkably long-lived positive photoconductivity with a lifetime extending 1 ns (without decay in 1 ns for some cases, see statistics for the measurements of 7 different sample areas in **Section S3** in SI). While the short-lived negative contribution can be attributed to the hot state of the graphene electronic system, the long-lived positive contribution appears only in the heterostructure. We assign this positive photoconductivity to an interfacial charge transfer process across g-WS$_2$ interfaces (which is supported by TA measurements in **Figure 3** below). The positive, long-lived photoconductivity is the result of photodoping: following photoexcitation, electrons are first transferred from graphene to WS$_2$ on a sub-ps timescale and subsequently get trapped at interfacial states where they remain for over 1 ns (see the detailed discussions in the last section of the report). For the initially p-doped graphene layer, extracting electrons from graphene results in a shift of Fermi level further away from the Dirac point, and thus an increase of conductivity in graphene.

To confirm such a Fermi level downshift following excitation and electron transfer from graphene to WS$_2$, we analyze the frequency dependence of the complex THz photoconductivity by THz-TDS (see details in **Section S2** in SI). The pump-probe delay ($\tau$) is chosen at ~100 ps to avoid the intrinsic hot carrier state in graphene, which has a lifetime of ~10 ps. As shown in **Figure 2B**, we find the frequency-resolved THz photoconductivity can be well-fitted with the Drude model, indicating that free carriers dominate the THz photoconductivity, with a scattering time of 69 fs (see details in **Section S2** in SI). Based on the fit, $E_F$ of graphene at $\tau = $ 100 ps following 1.55 eV excitation is 0.14 eV below the Dirac point (equivalent to $N = 1.1 \times 10^{12}$ cm$^{-2}$), indicating a 30-40 meV downshift of $E_F$ in graphene following electron transfer from graphene to WS$_2$ (given the initial $E_F$ of 0.1-0.11 eV).

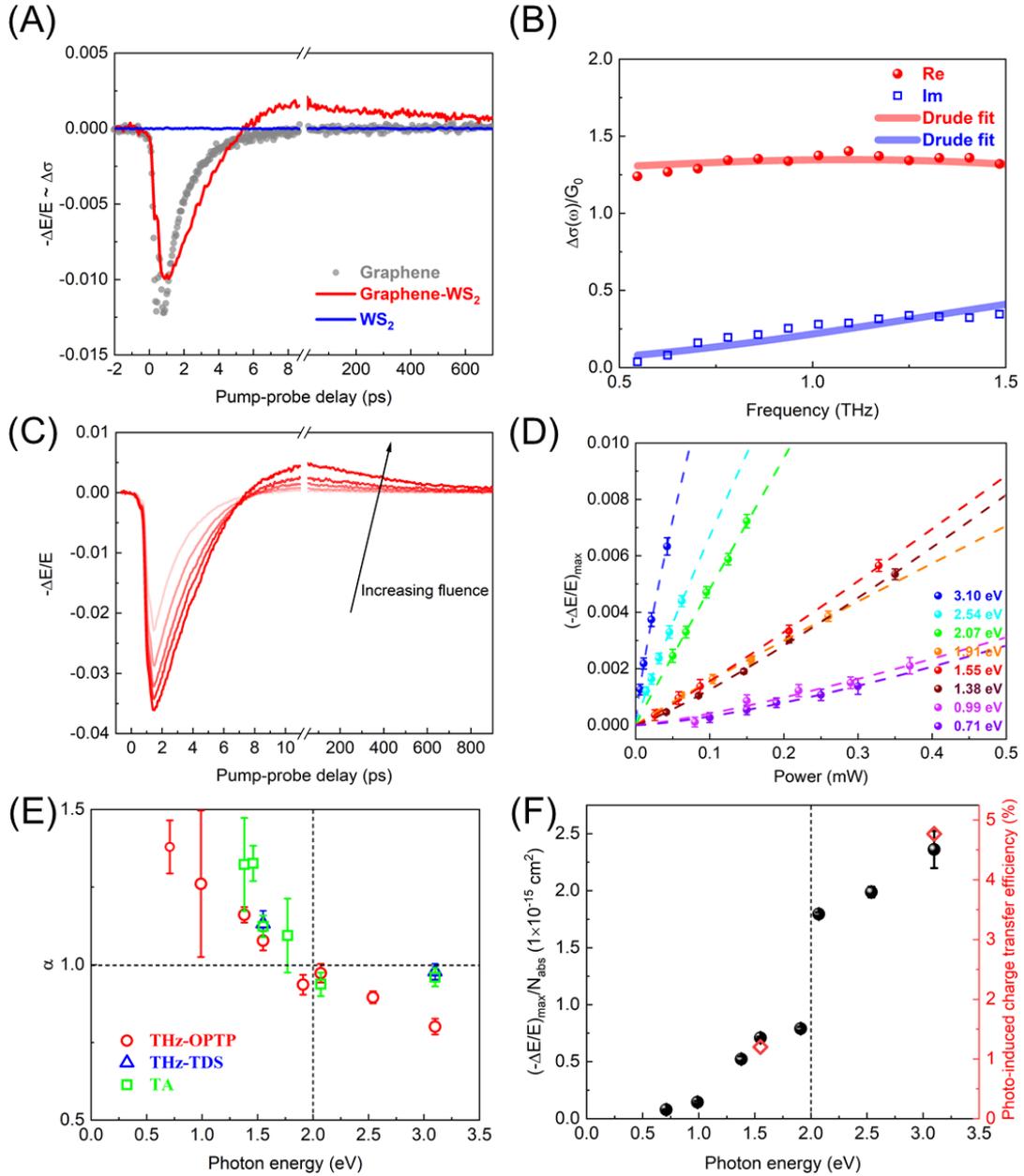

**Figure 2.** Investigation of ET at g-WS$_2$ interfaces by THz spectroscopy. (A) OPTP carrier dynamics for graphene (grey circle points), WS$_2$ (blue line), and the g-WS$_2$ heterostructure (red line). Samples are measured with 1.55 eV excitation under dry N$_2$ purging conditions. The absorbed photon densities are $3.8 \times 10^{12}$ cm$^{-2}$, $3.7 \times 10^{12}$ cm$^{-2}$, and $0.8 \times 10^{12}$ cm$^{-2}$ for WS$_2$, g-WS$_2$ heterostructure and graphene, respectively. (B) Complex photoconductivity for g-WS$_2$ heterostructure at a pump-probe delay of ~100 ps under 1.55 eV excitation with an absorbed photon density of $2.6 \times 10^{13}$ cm$^{-2}$. Red and blue lines represent the Drude fit for the real and imaginary part of the complex photoconductivity, respectively. We use the conductivity unit in quantum conductance $G_0 = 2e^2/h$; (C) Fluence-dependent ET dynamics at g-WS$_2$ interfaces. Samples are measured using 1.38 eV excitation with a series of absorbed photon densities, from $6.6 \times 10^{11}$ to $5.6 \times 10^{12}$ cm$^{-2}$. (D) Summary of the fluence-dependent photoconductivity maximum of $-\Delta E/E$ (i.e., $(-\Delta E/E)_{max}$) for different photon excitation

energies, from $h\nu$ = 0.71 to 3.10 eV. The data are described by a simple power law, namely $(-\Delta E/E)_{max} = A \cdot P^{\alpha}$ ($\alpha \geq 0$), as described in the main text; (E) Photon energy-dependent power index $\alpha$ obtained by THz-OPTP, THz-TDS and TA methods; (F) Quantification of photon energy-dependent ET efficiency in the heterostructure. The black dots represent the "relative" ET efficiency (the left Y axis), while the red diamonds stand for the "absolute" ET efficiency (the right Y axis) as defined in the paper. In (E-F), the A-exciton resonance energy of monolayer $WS_2$ (2 eV) is marked as a vertical dashed line.

*Electron transfer rate and mechanism at g-WS$_2$ interfaces*. Here we first focus on the nature of interfacial ET at g-WS$_2$ interfaces and quantifying its efficiency, before discussing the mechanism underlying the long-lived charge separation at g-WS$_2$ interfaces. We estimate the energy barrier for ET to be ~ 0.8 eV, by taking into account the energy difference between WS$_2$ conduction band minimum (CBM) and the Fermi level in graphene, as well as corrections including the bandgap reduction in WS$_2$ due to dielectric screening of graphene (*41*), and shifts in energetics associated with ET (see details in **Section S4** in SI). We found that the estimated energy barrier for ET is in line with previous studies (*23*). Given this large barrier and the fact that only graphene is excited by 1.55 eV pump photons, (thermalized or non-thermalized) hot carriers are necessarily involved in the interfacial charge transfer process at g-WS$_2$ interfaces. For most spectroscopic studies up to now, the optical excitation is limited to either below (*23*, *24*) or only at the A-exciton resonance of TMDCs (*25*). To investigate if and how the excitation energy or the initial charge configuration affects ET channels and mechanisms, we monitor the ET dynamics at various photoexcitation fluences and with a broad range of photon energies ($h\nu$) from 0.7 eV (which only excites graphene) to 3.1 eV (which excites both graphene and WS$_2$), across the A-exciton resonance (2 eV) of WS$_2$.

In **Figure 2C**, we present typical fluence-dependent ET dynamics for g-WS$_2$ heterostructure following 1.38 eV excitation. The long-lived positive photoconductivity increases with increasing absorbed photon density. To qualitatively describe the ET efficiency under different excitation conditions, we extract the maximum positive photoconductivity ($\Delta\sigma_{max} \sim (-\Delta E/E)_{max}$) for different pump energies and fluences, as shown in **Figure 2D**. For the pump photon energies and fluences employed, we always observe positive long-lived photoconductivity. This indicates that, for all cases, charge transfer involves electron injection from graphene to $WS_2$ (or equivalently, hole transfer from $WS_2$ to graphene). Intriguingly, for the lowest photoexcitation energy employed in this study (0.7 eV), the excess energy ($E_{ex}=h\nu/2-E_F$) in the photogenerated non-thermalized hot electrons is only ~ 0.45 eV, which is way below the energy barrier for ET at g-$WS_2$ interfaces (~0.8 eV). The fact that we nonetheless observe a positive photoconductivity indicates that ET takes place following thermalization via so-called photo-thermionic emission (*21*). This photo-thermionic emission scheme relies on very high electron temperatures in graphene reached for typical incident excitation powers. Sufficiently hot electrons in the valence band can be heated across the Dirac point into the conduction band. Following this interband heating process (*38*), the thermalized hot electrons with sufficient energy over the energy barrier at g-$WS_2$ interfaces can be injected into $WS_2$. We calculate the electron temperature ($T_e$) in graphene as a function of incident fluence based on the photo-induced THz response of graphene (see details in **Section S5** in SI). As shown in **Figure S3(A)**, $T_e$ can reach 1500–3000 K over our fluence range. **Figure S3(C)** shows the resultant hot electron distribution following the thermalization. A substantial fraction of thermalized hot electrons (~22.2 % with $T_e$ of 2700 K) reaches energies in excess of the conduction band in $WS_2$, in good agreement with our proposed photo-thermionic emission scheme. We conduct a similar analysis for thermalized hot holes in the valence band in graphene. We observe a much smaller fraction (~1.6 % with $T_e$ of 2700 K) of thermalized hot holes with energy above the energy barrier for hole injection than that of thermalized hot electrons in our system (see **Figure S3(D)**), due to the larger energy separation between the Fermi level in graphene and the valence band in $WS_2$ (~ 1.3 eV). This result can explain our experimental observation of the injection of thermalized hot electrons rather than holes from graphene to $WS_2$ in the study.

In principle, the photoconductivity caused by the photo-thermionic emission should exhibit a superlinear dependence on the pump fluence (*21*, *22*), see also the simulation result shown in **Figure S3(E)**. We therefore describe the pump fluence dependent maximum positive photoconductivity with $P^\alpha$ as shown in **Figure 2D**, where $P$ is the pump power, and $\alpha$ is the power index ($\alpha > 1$ for superlinear dependence). In **Figure 2E**, we show the extracted values of $\alpha_{THz\text{-}OPTP}$ for different photon excitation energies. We find that $\alpha_{THz\text{-}OPTP}$ is bigger than 1 when the photon excitation energy is below the A-exciton resonance of WS$_2$. With increasing photon excitation energy, $\alpha_{THz\text{-}OPTP}$ gradually decreases and undergoes a transition around the A-exciton resonance of WS$_2$, from above 1 ($\alpha$ ~1.4 at 0.7 eV excitation, superlinear) to slightly below 1 ($\alpha$ ~0.9 for $h\nu > 2\ eV$, sublinear). It is worth noting that in previous TA studies (*23*, *24*), a linear fluence dependence (i.e., $\alpha = 1$) of the signal was reported, following sub-A-exciton excitation. Therefore, to further verify our THz results, we have repeated the fluence dependent measurements for the same sample employing other complementary methods, including THz-TDS and TA spectroscopy (see details of our TA study in the next section). The power indexes $\alpha$ obtained from these three methods are found to follow nearly the same trend as shown in **Figure 2E** (experimental data, a detailed analysis and discussions can be found in **Section S6** in SI). This result shows that the change in $\alpha$ is independent of the employed spectroscopic tools.

We attribute this superlinear-to-sublinear pump fluence dependence of the ET dynamics to a transition between two distinct ET regimes: HET via photo-thermionic emission for sub-A-exciton excitation (see **Figure 1C**), and direct hole transfer (DHT) for above-A-exciton excitation. The DHT process involves the interfacial recombination of holes in the valence band of WS$_2$ with valence band electrons in graphene (see **Figure 1D**). A slightly sub-linear feature in our study can be understood as follows: with increasing fluence, many-body effects, for instance, exciton-exciton annihilation in TMDCs (*42*), can play a critical role in the charge carrier dynamics on the sub-ps to ps time scale. Such many-body effects can result in a decrease of the hole density in TMDCs, which reduces the charge transfer efficiency at high fluence. Note that in the above-A-exciton excitation regime, hot electrons in graphene, in spite of having higher energy than those generated by sub-A-exciton excitation (for a given absorbed power), contribute little to ET. This is mainly due to the much weaker absorption in graphene in comparison to that in WS$_2$. On top of that, we discuss other ET pathways including photo-thermionic emission, and hole transfer from WS$_2$ to the photoexcited hot electrons in graphene in the above-A exciton excitation regime (see the extended discussions in **Section S7** in SI), and argue that they cannot compete with DHT process.

It is worth commenting that our results and proposed microscopic model agree well with a recent study by time-resolved angle-resolved photoemission spectroscopy (tr-ARPES). In the study (*25*), an ultrafast (sub-200 fs) loss of the valence band electrons in graphene has been directly observed in the same heterostructure by resonantly exciting the A-exciton resonance of monolayer WS$_2$. Furthermore, for the photocurrent generation in g-based heterostructures (*21-22*), the photo-thermionic emission is proposed to govern the photocurrent generation for low photoexcitation energy, whereas direct interlayer tunneling becomes the dominant mechanism for high photoexcitation energy. Our results here are also in line with these macroscopic device results and further provide macroscopic mechanisms for the interfacial ET processes. Finally, despite the clear difference in ET mechanisms between the two regimes discussed, we note that the resultant charge configurations at the interfaces are indistinguishable with electrons in the excited state of WS$_2$ and holes in the valence band of graphene.

To further confirm the proposed two ET regimes, we quantify their efficiency at different excitation energies. Here we utilize two approaches to determine the interfacial ET efficiency based on THz-TDS and OPTP results: (a) we define the absolute ET efficiency as: $\eta = N_{ET}(hv)/N_{abs}$, where $N_{ET}(hv)$ is the number of electrons injected from graphene to WS$_2$ at excitation energy $hv$, and $N_{abs}$ represents the absorbed photon density in the donor, i.e., graphene for the photo-thermionic emission regime and WS$_2$ for the DHT regime. $N_{ET}$ and $N_{abs}$ are experimentally accessible by THz-TDS (see details in **Section S2** in SI) and absorption measurements (see details in **Section S1** in SI), respectively; (b) based on the OPTP results shown in **Figure 2D**, we can directly compare the maximum positive photoconductivity at a fixed absorbed photon density (~ of $5 \times 10^{12}$ cm$^{-2}$). At the pump-probe delay where the positive photoconductivity is maximum (~15 ps), charge carriers in our system reach quasi-equilibrium following ultrafast ET, so that the carrier mobility $\mu$ of graphene is nearly constant over time. As the photoconductivity $\Delta\sigma$ is proportional to the products of $\mu$ and $N_{ET}(hv)$, $\Delta\sigma/N_{abs}$ at the maximum positive photoconductivity reflects the relative change in the ET efficiency under different photon excitation energies.

In **Figure 2F**, we plot the ET efficiencies obtained by (a) THz-TDS (as right Y-axis) and (b) OPTP (as left Y-axis) as a function of photon excitation energy. For 1.55 eV and 3.10 eV excitations, where both methods are applied, the ET efficiencies evaluated by these two methods agree perfectly, validating our proposed methods. We observe that the ET efficiency shows a transition across the A-exciton resonance of WS$_2$: below it, the ET efficiency is relatively inefficient with a quantum yield of < 1%; above it, the ET efficiency increases with the increasing excitation energy and reaches ~ 5% for 3.10 eV excitation. One of the origins for the much-enhanced ET efficiency in the DHT regime in comparison to that for the photo-thermionic emission regime, could be attributed to the density of electrons in graphene for ET: in the photo-thermionic emission regime, only a limited number (on the order of $10^{11}$~$10^{12}$ cm$^{-2}$) of electrons with sufficiently high energy can be emitted over the energy barrier, while in the DHT regime, the valence band electrons in graphene, with a much higher density (on the order of ~$10^{14}$ cm$^{-2}$), can efficiently recombine with the photogenerated holes in the valence band of WS$_2$.

Our data here provides direct and strong evidence for the transition of the ET mechanism from HET (via photo-thermionic emission when only graphene is excited) to DHT between the valence bands from WS$_2$ to graphene (in the above-A-exciton excitation regime). Finally, in the sub-A-exciton excitation regime, while it is clear that HET via photo-thermionic emission takes place ($\alpha_{THz\text{-}OPTP} > 1$ for $hv < 2.0$ eV), it remains ambiguous if other HET pathways are possible. To check this aspect, we re-normalize the maximum positive photoconductivity to the incident power ($P$, rather than $N_{abs}$ as shown in **Figure 2F**) and replot it *vs.* the pump photon energy. As the electron temperature in graphene is independent on the pump photon energy for a fixed incident power, the HET efficiency via pure photo-thermionic emission will be constant for different pump photon energies in such a plot (*21, 22*). As shown in **Figure S6**, for $hv < 1$ eV, the HET efficiency is nearly constant, in line with photo-thermionic emission; for $1.4 < hv < 2$ eV, we observe an increase of the HET efficiency. This indicates that in the sub-A-exciton excitation regime, with sufficiently high pump photon energy, other possible ET routes may also contribute to the ET process. One of the plausible scenarios is ET involving non-thermalized hot electrons (on top of photo-thermionic emission). Knowing the energy difference between the $WS_2$ CBM and graphene's Dirac point (~ 0.7 eV), the minimum pump energy required for non-thermalized hot electron injection is $2 \times 0.7$ eV $= 1.4$ eV, in line with the experimental observation. However, further studies are needed to rule out other possible ET mechanisms, such as DHT at graphene and small domains of multi-layer $WS_2$ (with absorption onset around 1.3 eV) (*21*).

***ET dynamics and interfacial charge separation time probed by transient absorption spectroscopy.*** Finally, to provide a direct spectroscopic signature of electron injection from graphene to $WS_2$, we have performed complementary measurements using TA spectroscopy. Along with providing direct evidence for ET across the heterostructures, employing TA spectroscopy is further motivated by noticing a large discrepancy between the charge separation lifetime obtained by our THz results (beyond 1 ns) and previous TA results (reported to be ~1 ps). By probing the corresponding excited state absorption changes with a pump-probe scheme, the charge occupation dynamics in $WS_2$ can be directly obtained with a sub-ps time resolution. As shown in **Figure 3A**, we observe no photobleaching (PB) signal when exciting the individual monolayer $WS_2$ with 1.55 eV pulses (below its A-exciton resonance), in line with our THz results (**Figure 2A**, blue line). This result also indicates that the pump fluence used in this study is sufficiently low to avoid significant two-photon absorption in $WS_2$. A small, short-lived differential-like dynamic (with a lifetime of ~150 fs as shown in **Figure**

**3C**) is noticed and may be attributed to the coherent artifact, predominantly as a result of the optical stark effect, which has been widely reported previously (*43*). In contrast, for the g-WS$_2$ heterostructure, we observe two PB signals at both the A- and B-exciton resonances of monolayer WS$_2$ (2.0 eV and 2.4 eV, respectively) for the same excitation energy and similar pump fluence, as shown in **Figure 3B**. For monolayer WS$_2$, A- and B- exciton resonances originate from two spin-split valence bands at the *K* point (*29, 30*). The simultaneous occurrence of PB at both A- and B-exciton resonances provides an additional evidence of the injection of (hot) electrons rather than holes from graphene to WS$_2$ following sub-A-exciton excitation, which is in line with THz data.

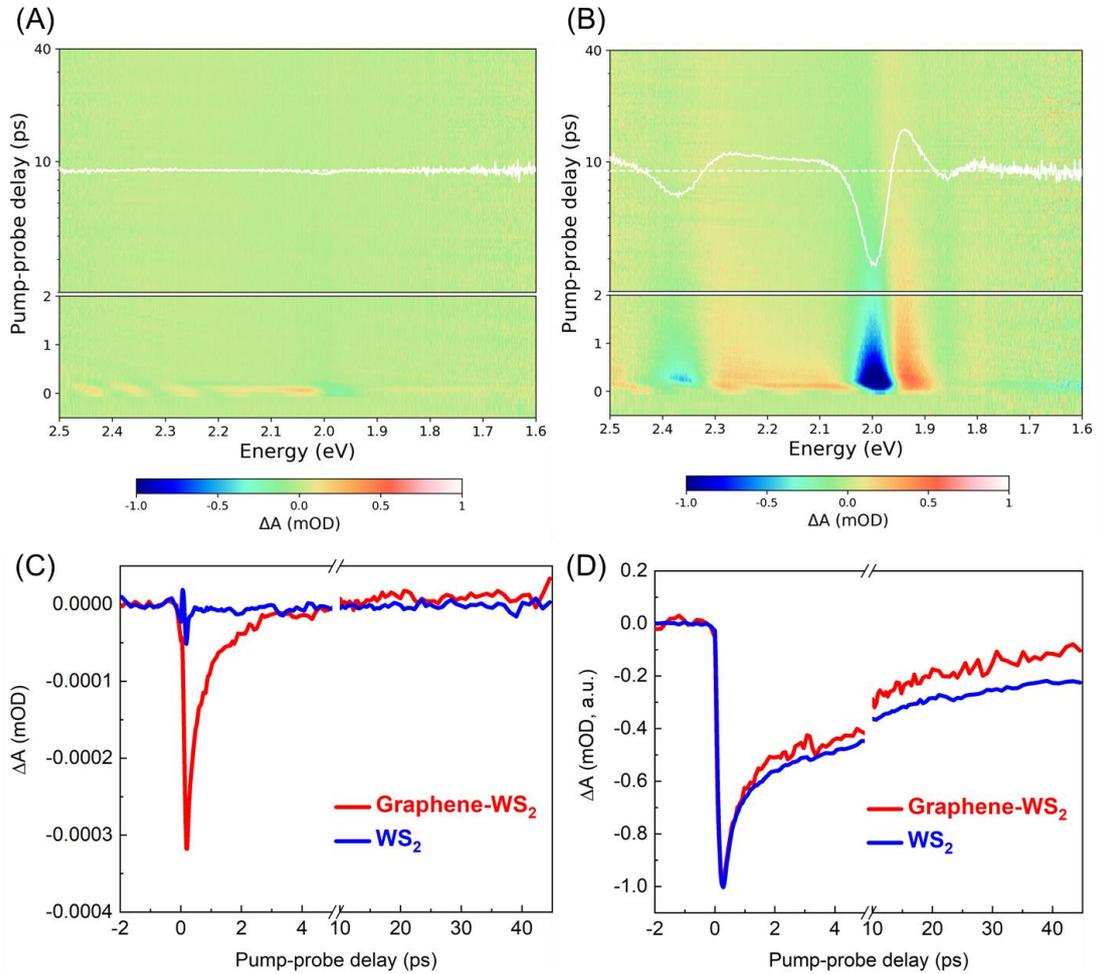

**Figure 3.** Investigation of ET at g-WS$_2$ interfaces by TA spectroscopy. 2D TA color map for (A) monolayer WS$_2$, and (B) g-WS$_2$ heterostructure following 1.55 eV excitation with a pump fluence of 670 μW and 620 μW, respectively. The white lines in Figure 3A and Figure 3B represent the spectral slices averaged between 0.5 and 1 ps. (C) TA carrier dynamics averaged over the A-exciton resonance (between 1.88 eV and 2.07 eV) for g-WS$_2$ heterostructure (red

line) and monolayer WS$_2$ (blue line) following 1.55 eV excitation with a pump fluence of 670 µW and 620 µW, respectively. (D) TA carrier dynamics averaged over the A-exciton resonance (between 1.88 eV and 2.07 eV) for g-WS$_2$ heterostructure (red line) and monolayer WS$_2$ (blue line) following 3.10 eV excitation with a pump fluence of 248 µW and 250 µW, respectively.

To quantify the ET rate and lifetime of the injected electrons in WS$_2$, we show the PB dynamics averaged over the A-exciton resonance of WS$_2$ in **Figure 3C** (red line). The rise time of the PB, corresponding to the ET process from graphene to the first excited state of WS$_2$, is found to be less than 150 fs. In principle, for a given pump photon density, the ET rate is expected to increase by increasing the pump photon energy. However, we observe no difference in the rising dynamics at different excitation energies (see **Figure S7**) from our TA results, indicating the dynamics are limited by the time resolution of the setup (~ 150 fs). Indeed, the ET rate has previously been reported to be < 25 fs (*23*). Furthermore, we find that the injected electrons in the excited states in WS$_2$ decay within 1 ps, in line with previous TA measurements (*23, 24*). In these studies, the ultrafast depopulation process was attributed to back electron transfer from WS$_2$ to graphene. Surprisingly, the long-lived photoconductivity measured by THz spectroscopy on the sample with the same configuration shows that the injected electrons in WS$_2$ do not recombine with the remaining holes in graphene immediately. The large difference in the measured carrier lifetime in the donor (graphene) and the acceptor (WS$_2$) thus suggests a new recombination pathway.

***Interfacial photogating effect in g-WS$_2$ heterostructure is supported by interfacial states.*** Here we attribute the long-lived (over ~ 1 ns) photoconductivity in graphene (by THz spectroscopy) to an interfacial photogating mechanism mediated by interfacial states (e.g., defects) at g-WS$_2$ interfaces. In such a scheme, following photoexcitation, electrons in WS$_2$, either directly generated by excitation (with *hυ* > 2eV) or injected from graphene to the conduction band of WS$_2$ (for *hυ* < 2eV), are subsequently trapped (within ~1 ps) to long-lived interfacial states. The relatively long-lived "trapped" electrons can effectively gate graphene, leading to an efficient modification of the carrier density (thus the Fermi level) in graphene (observed by THz spectroscopy). Intriguingly, following the fast ~1 ps PB decay, the TA dynamics is dominated by a much longer-lived, spectral shift (with the differential-like feature, see the dynamics and the spectral slices in **Figure 3B**). This feature can be understood by transient electrical field induced Stark effect, following ultrafast ET and trapping at the interface. Such spectral shift due to the local electrical field built at the interfaces has previously been reported in heterostructures consisting of semiconducting quantum dots (*44*).

**Discussion**

Now we briefly discuss the possible origin of the "interfacial" states. First, as the energies of these states lie between the conduction band of $WS_2$ and the Fermi level of graphene, it seems reasonable to assume that they originate from defect states present in the $WS_2$ layer, or hybridized states at interfaces. To test this hypothesis and shed light on the nature of these states, we have studied and compared the TA dynamics of monolayer $WS_2$ and g-$WS_2$ heterostructure following 3.10 eV excitation. As shown in **Figure 3D**, we observe very similar decay dynamics for both samples (especially for the fast initial decay). This result strongly suggests a universal recombination pathway at the first 1-2 ps, very likely due to trapping in the defect states in $WS_2$. Note that this assumption is in line with previous THz and TA dynamics reported in monolayer TMDCs, where the fast, ~ 1 ps decay is also assigned to charge trapping (*45–47*).

According to literature, we speculate that the defects states could originate from sulfur vacancies, which are ubiquitously present in CVD-$WS_2$. For instance, a recent report combining ab initio GW calculations and scanning tunneling spectroscopic studies has shown that sulfur vacancies in $WS_2$ can generate two types of unoccupied in-gap states, which are located ~0.5 and 0.7 eV below the conduction band of $WS_2$, respectively (*48*), with at least one of them lying between the Fermi level of graphene and the conduction band of monolayer $WS_2$. These empty defect states can serve as the "intermediate" trap sites, which can electrostatically gate the graphene layer. Further studies employing photo-electrochemical methods (*49–51*) to unveil the energetics and the nature of the involved defects at the g-WS$_2$ interfaces could shed light on this issue.

We unveil new injection and recombination pathways in g-TMDCs, and demonstrate the beneficial effect of interfacial defect states for promoting a long interfacial charge separation time and thus an efficient photogating phenomenon in g-based vdW structures. Understanding such injection and recombination pathways is important for both fundamental studies and optoelectronic applications. For instance, the long-lived photogating effect reported here (mediated by ultrafast interfacial ET and interfacial trapping process) reconciles the reported very short charge separation time at g-TMDCs interfaces (*23*, *24*) and however very efficient photodetectors based on these structures (*5–7*). While the interfacial defects are beneficial in such circumstance, they may be troublesome for other applications, such as photovoltaics based on such vdW structures: the short lifetime of injected hot electrons in the conduction band of TMDCs imposes a limited time window for the efficient separation of charges towards electrodes. Passivation of the interfacial states or development of further extraction of injected hot electrons from TMDCs through an ultrafast, sub-ps channel is required.

**Conclusion**

In summary, we investigate the nonequilibrium hot carrier dynamics in g-WS$_2$ vdW heterostructures combining ultrafast THz and TA spectroscopy. We report a transition in both ET efficiency and mechanism by tuning the pump photon energies across the A-exciton resonance of WS$_2$. Upon excitation below the A-exciton resonance of WS$_2$, a relatively inefficiently HET via photo-thermionic emission governs the ET process and only thermalized hot electrons with sufficient energy can be injected into the excited states of WS$_2$. In contrast, we show that highly energetic non-thermalized hot electrons in graphene do not contribute to the ET process. Rather, a relatively efficient (up to ~5%) direct hole transfer (DHT) process occurs from the valence band of WS$_2$ to the valence band of graphene. Importantly, we show that the injected electrons only occupy the excited states of WS$_2$ for ~1 ps and then get trapped and stored at vdW interfaces (probed by TA spectroscopy). This results in a long-lived photogating effect in graphene over 1 ns (observed by THz spectroscopy). Our results here provide new insights into both the ET mechanism and recombination pathway at g-TMDCs vdW interfaces, which are critical to potential optoelectronic and energy-harvesting applications of g-TMDCs vdW heterostructures.

**Acknowledgments**

We thank Paniz Soltani, Aliaa S. Hassan, Zhaoyang Liu, Marco Ballabio, Alexander Tries and Mathias Kläui for fruitful discussions. S.F acknowledges the fellowship support from Chinese Scholarship Council (CSC). X.J. acknowledges the financial support by DFG through the Excellence Initiative by the Graduate School of Excellence Materials Science in Mainz (MAINZ) (GSC 266) and support from the Max Planck Graduate Center mit der Johannes Gutenberg-Universität Mainz (MPGC). A.J.H. acknowledges support from the European Research Council Horizon 2020 ERC Grant No. 678004 (Doping on Demand). ICN2 was supported by the Severo Ochoa program from Spanish MINECO (Grant No. SEV-2017-0706). K.J.T. acknowledges funding from the European Union's Horizon 2020 research and innovation programme under Grant Agreement No. 804349 (ERC StG CUHL), and financial support through the MAINZ Visiting Professorship.

**Author contributions**

H.I.W. conceived and supervised the project. S.F. conducted THz studies with help from X.J., J.X., X.Y., H.Z., W.Z, S.K.; I.F. conducted TA measurements, analyzed the data under the supervision of A.J.H.; X.J. and K.-J.T. modeled the electron temperature in graphene based on THz conductivity; all authors contributed to data interpretation; S.F. and H.I.W. wrote the paper with input from all authors.

**Competing interests**

The authors declare no competing interests.

Supporting information for

# Long-Lived Charge Separation Following Pump-Energy Dependent Ultrafast Charge Transfer in Graphene/WS$_2$ Heterostructures

Shuai Fu[1], Indy du Fossé[2], Xiaoyu Jia[1], Jingyin Xu[1,3], Xiaoqing Yu[1], Heng Zhang[1], Wenhao Zheng[1], Sven Krasel[1], Zongping Chen[4], Zhiming M. Wang[3], Klaas-Jan Tielrooij[5], Mischa Bonn[1], Arjan J. Houtepen[2], Hai I. Wang[1*]

[1] *Max Planck Institute for Polymer Research, Ackermannweg 10, D-55128 Mainz, Germany*

[2] *Optoelectronic Materials Section, Faculty of Applied Sciences, Delft University of Technology, Van der Maasweg 9, 2629 HZ Delft, The Netherlands*

[3] *Institute of Fundamental and Frontier Sciences, University of Electronic Science and Technology of China, Chengdu, 610054 P. R. China*

[4] *School of Materials Science and Engineering, Zhejiang University, 38 Zheda Road Hangzhou 310027, China*

[5] *Catalan Institute of Nanoscience and Nanotechnology (ICN2), BIST and CSIC, Campus UAB, Bellaterra, 08193 Barcelona, Spain*

*E-mail: wanghai@mpip-mainz.mpg.de

**Content**

**Section S1.** Static optical characterizations of the g-WS$_2$ heterostructure

**Section S2.** Terahertz-time domain spectroscopy (THz-TDS)

**Section S3.** Recombination dynamics of the separated charge carriers following ultrafast HET from graphene to WS$_2$ at varied spots

**Section S4.** Estimation of the energy barrier for electron transfer and hole transfer

**Section S5.** Calculation for thermalized hot electron transfer (HET) and thermalized hot hole transfer (HHT) in the g-WS$_2$ heterostructure

**Section S6.** $α_{TA}$ and $α_{THz\text{-}TDS}$ at different photon excitation energies

**Section S7.** Discussion on ET pathways involving photoexcited hot electrons in graphene in the above-A-exciton excitation regime

**Section S8.** Photon energy-dependent maximum positive photoconductivity normalized to the incident power

**Section S9.** Normalized photon excitation energy dependent PB dynamics in the g-WS$_2$ heterostructure

**Section S10.** TA spectrum for the g-WS$_2$ heterostructure averaged at different pump probe delay times upon 1.55 eV excitation

## Section S1. Static optical characterizations of the g-WS₂ heterostructure

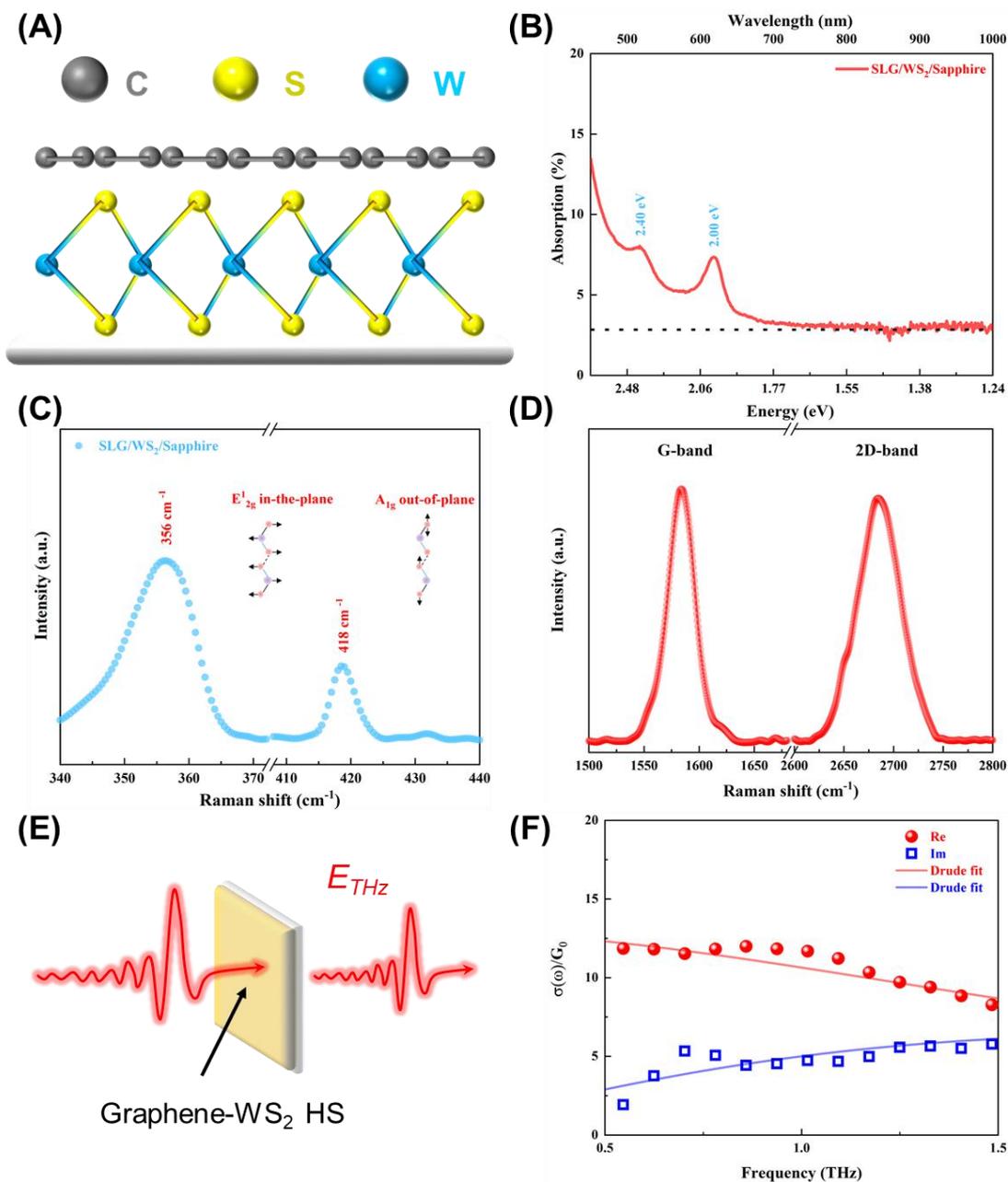

**Figure S1.** Optical characterizations of the g-WS₂ heterostructure. (A) Schematic of graphene-WS₂ heterostructure deposited on a sapphire substrate; (B) UV-vis absorption of the heterostructure; (C) Raman spectra recorded in the characteristic region of WS₂; (D) Raman spectra recorded in the characteristic region of graphene; (E) Illustration of terahertz time domain spectroscopy (THz-TDS); (F) Complex conductivity ($\sigma(\omega)$) of graphene in the unit of quantum conductance $G_0 = 2e^2/h$. The red and blue lines represent the Drude fits described in this paper.

**Figure S1(A)** depicts the graphene-WS$_2$ heterostructure used in this study, which is produced by transferring a chemical-vapor-deposited (CVD) graphene monolayer onto a CVD WS$_2$ monolayer supported on a sapphire substrate. **Figure S1(B)** shows the UV-vis absorption spectra of the heterostructure and has been discussed in the main text. **Figure S1(C)** and **Figure S1(D)** present the characteristic Raman regions for WS$_2$ and graphene, respectively. The characteristic in-plane ($E_{2g}^{-1}$) and out-of-plane ($A_{1g}^{-1}$) vibrational modes for WS$_2$ are found to be at 356 cm$^{-1}$ and 418 cm$^{-1}$, respectively, in perfect agreement to those for monolayer WS$_2$ (*1*). Raman modes centered at 1584 cm$^{-1}$ and 2684 cm$^{-1}$ are assigned to the well known G-band and 2D-band in graphene. In addition, as the frequency of G-band ($w_G$) is highly sensitive to the Fermi level ($E_F$) in graphene (*2, 3*), one can evaluate the Fermi level in graphene following an empirical equation as shown in Eq. (S1) (*2, 4*):

$$w_G - 1580 cm^{-1} = (42 cm^{-1} eV)|E_F| \quad (S1)$$

Based on this method, $E_F$ is found to be ~ 0.11 eV below the Dirac point (given the p-doped nature of graphene from electrical measurements).

## Section S2. Terahertz-time domain spectroscopy (THz-TDS)

### *(A) Quantifying the initial carrier density in graphene by THz-TDS*

We have employed terahertz time domain spectroscopy (THz-TDS) to characterize electrical properties of the heterostructure, without photoexcitation, as depicted in **Figure S1(E)**. To this end, we record the fields of THz waveform transmitted through the WS$_2$/sapphire substrate configurations in the time domain with and without graphene on the top, i.e., *E(t)* (for graphene/WS$_2$/sapphire*)* *and* *E$_0$(t)* (for WS$_2$/sapphire*)*, respectively. The obtained time-dependent THz fields are further converted into the frequency domain by Fourier transform as *E(ω) and E$_0$(ω)*. The difference between *E(ω) and E$_0$(ω)* is due to the presence of graphene. Based on the standard thin-film approximation (*5*, *6*), we determine the frequency-dependent complex sheet conductivity *σ(ω)* of graphene using Eq. (S2),

$$\frac{E(w)}{E_0(w)} = \frac{n+1}{n+1+z_0\sigma(w)} \quad (S2)$$

$$\sigma(w) = \frac{D}{\pi}\frac{1}{(\Gamma - i\omega)} \quad (S3)$$

where $Z_0 = 377\ \Omega$ is the impedance of free space, and n = 1.77 is the THz refractive index of the sapphire substrate. As shown in **Figure S1(F)**, the obtained complex conductivity *σ(ω)* exhibits a free-carrier response following Drude behaviors (*7–9*) in our THz spectral window. By fitting the obtained complex conductivity *σ(ω)* with a simple Drude model as shown in Eq. (S3), we extract two parameters, i.e., Drude weight *D* and scattering time $\tau$. For conductivity stems from free charge carriers in graphene with carrier density *N*, the Drude weight is given by Eq. (S4) (*8–10*),

$$D = (v_F e^2/\hbar)(\pi N)^{1/2} \quad (S4)$$

where $v_F$ is the Fermi velocity (1.1×10$^6$ m/s), $\hbar$ is the reduced Planck constant. By correlating the obtained Drude weight *D* with the carrier density *N* using Eq. (S4), we estimate the carrier density in graphene as $7.9 \times 10^{11}$ cm$^{-2}$. This corresponds to a Fermi level of ~0.1 eV according to Eq. (S5).

$$|E_F| = \hbar v_F (\pi|N|)^{1/2} \quad (S5)$$

The estimated Fermi levels of graphene by THz-TDS and Raman spectroscopy are in good agreement with each other.

***(B) Quantifying the pump-induced carrier density changes in graphene by THz-TDS***

In order to quantify the pump-induced carrier density change in graphene following optical excitation, we perform THz-TDS measurements at pump-probe delay τ = 100 ps (τ >> 10 ps to avoid the hot carrier effect in graphene). As the conductivity of graphene at τ = 100 ps, i.e., $\sigma_\tau(\omega)$, also follows Drude response, the pump-induced conductivity change can be described by the differential change of two Drude response terms with their own Drude weights and scattering times as shown in Eq. (S6).

$$\Delta\sigma_\tau(w) = \sigma_\tau(w) - \sigma(w) = \frac{D_\tau}{\pi}\frac{1}{(\Gamma_\tau - iw)} - \frac{D}{\pi}\frac{1}{(\Gamma - iw)} \quad (S6)$$

where $D$ and $G$ are the initial Drude weight and scattering rate (obtained from the Drude fit as shown in **Figure S1(F)**), $D_t$ and $G_t$ are the Drude weight and scattering rate at τ = 100 ps. The carrier density $N_t$ can be calculated out by Eq. (S4). Therefore, the photo-induced carrier density change in graphene at τ = 100 ps can be estimated as $\Delta N_\tau = N_\tau - N$.

**Section S3. Recombination dynamics of the separated charge carriers following ultrafast HET from graphene to WS$_2$ at varied spots**

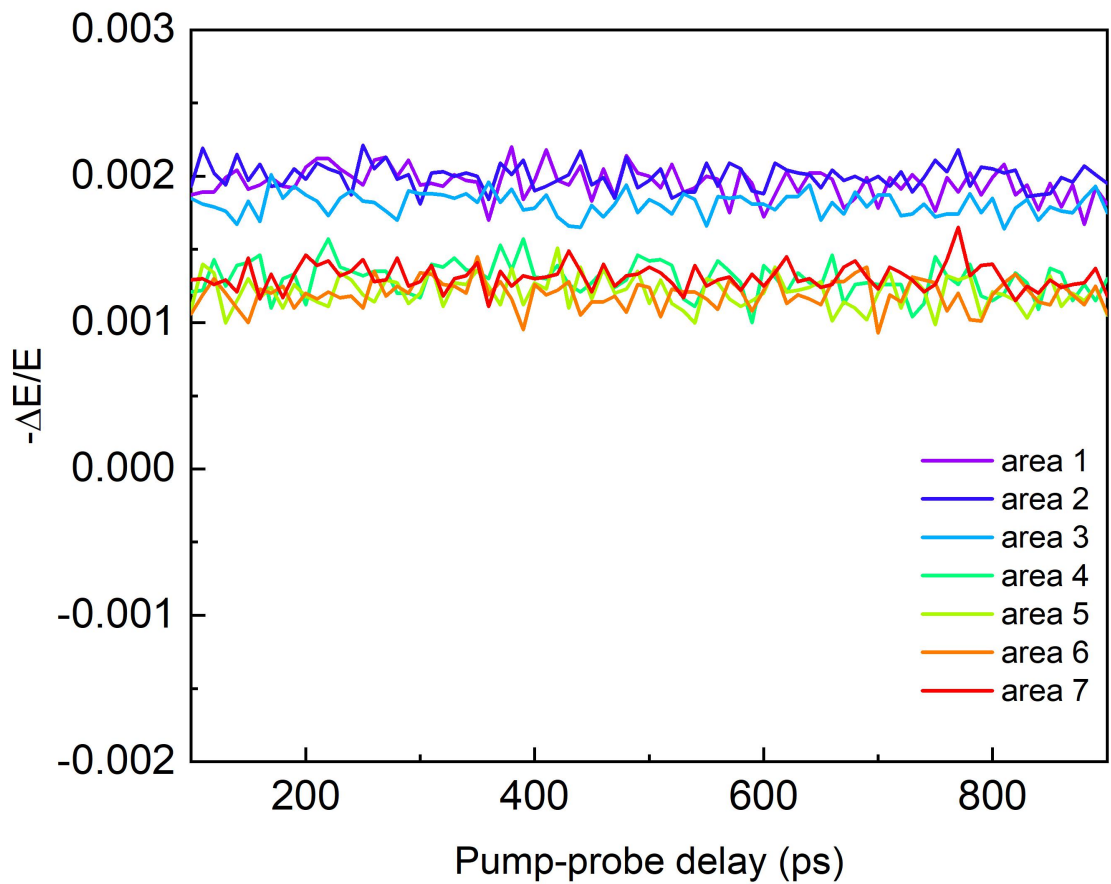

**Figure S2.** Recombination dynamics of the separated charge carriers following ultrafast HET from graphene to WS$_2$ at 7 different spots. Measurements are performed under 1.55 eV excitation with a pump fluence of 188 μW.

## Section S4. Estimation of the energy barrier for electron transfer and hole transfer

Based on previous tr-ARPES studies (*11*, *12*), the energy difference between the Dirac point in graphene and the valence band maximum (VBM) in monolayer $WS_2$ is ~ 1.5 eV. The electronic or quasi-particle bandgap of $WS_2$ is reported to be ~ 2.3 eV (*13*). This means, in our study, the "static" energy difference between the conduction band minimum (CBM) in $WS_2$ and the Fermi level in graphene is 0.9 eV (2.3 eV - 1.5 eV - (-0.1 eV) = 0.9 eV, given the initial Fermi level in graphene is -0.1 eV in our study).

Now we consider other contributions associated with (1) bandgap normalization and (2) shift of energetics due to charging and "coulomb effect" following the charge separation at g-$WS_2$ interfaces:

For (1), the bandgap of $WS_2$ on graphene is reported to go down ~0.2 eV due to the strong screening effect of graphene and substrate. As the effective mass of electron and hole in $WS_2$ is found to be very close (0.44 $m_0$ for electron and 0.45 $m_0$ for hole) (*14*), such screening effect leads to a reduction of ~0.1 eV for both the CBM and the VBM in $WS_2$, and thus the energy barrier for electron transfer and hole transfer;

For (2): following ET from graphene to $WS_2$, the presence of charge carriers in $WS_2$ will ""push up" the CBM in $WS_2$, while the charging in graphene will lower the Dirac point. The upshifted CBM in $WS_2$ and the downshifted Dirac point in graphene will increase the energy barrier for ET. On the other hand, the interlayer coulomb interaction will lead to a reduction in energetics by forming the "interfacial charge transfer state". Direct qualification of these three contributions individually is challenging, but we can estimate the total shift from a recent tr-ARPES paper (*11*): with a total transferred charge density of $5 \times 10^{12}$ $cm^{-2}$, the shift in energetics in $WS_2$ is ~ +90 meV (upshift), while the shift of the valence band in graphene is ~ -50 meV (downshift). This leads to an increase of the energy barrier for ET of 140 meV. For our case, the transferred charge density is ~$1.7 \times 10^{11}$ $cm^{-2}$. Assuming the shift in energetics scales linearly with the transferred charge density, this leads to an increase of the energy barrier for ET of ~ 5 meV (140 meV × ($1.7 \times 10^{11}$ $cm^{-2}$/$5 \times 10^{12}$ $cm^{-2}$) ≈ 5 meV).

Therefore, by taking into account the energy difference between the CBM in $WS_2$ and the Fermi level in graphene, as well as corrections including the bandgap reduction in $WS_2$ due to dielectric screening of graphene, and shifts in energetics associated with ET, we estimate the energy barrier for ET to be ~ 0.8 eV. (0.9 eV -0.1 eV + 0.005 eV ≈ 0.8 eV). Correspondingly, we estimate the energy barrier for hole transfer to be ~ 1.3 eV (2.1 eV - 0.8 eV = 1.3 eV).

**Section S5. Calculation for thermalized hot electron transfer (HET) and thermalized hot hole transfer (HHT) in the g-WS$_2$ heterostructure**

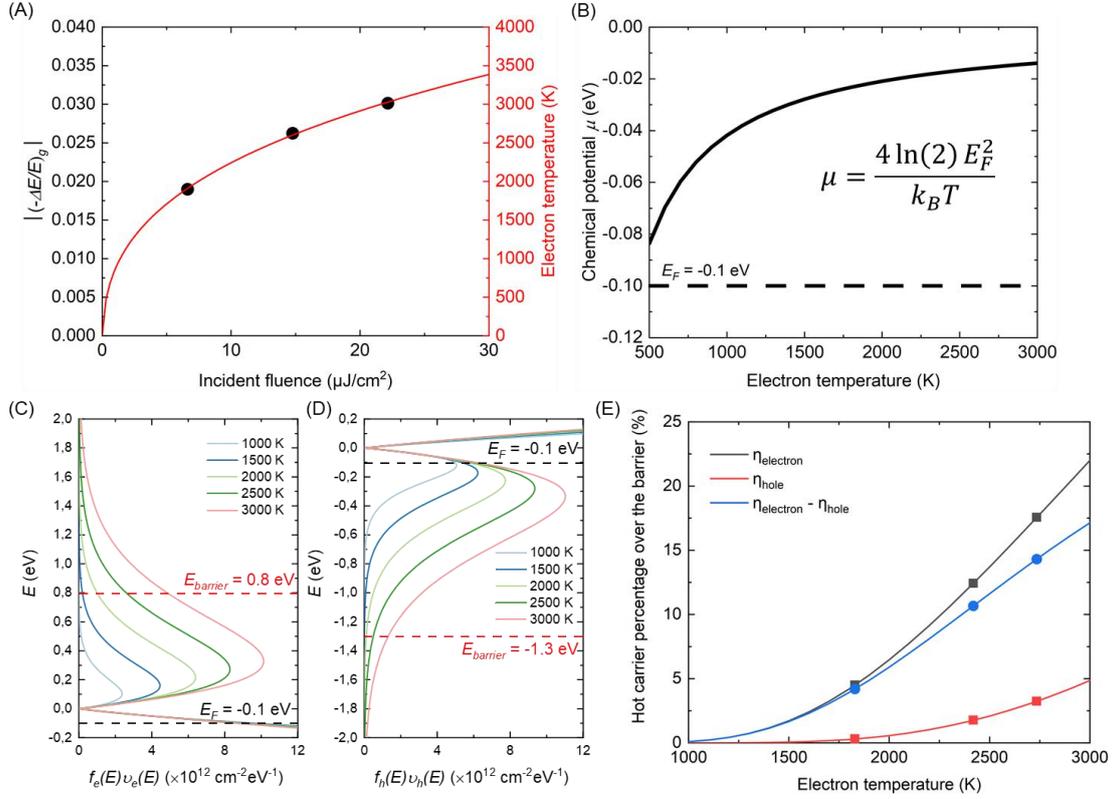

**Figure S3.** Calculation for thermalized hot electron transfer (HET) and thermalized hot hole transfer (HHT) in the g-WS$_2$ heterostructure. (A) THz peak conductivities (left Y axis, absolute value) and the corresponding $T_e$ (right Y axis) for bare graphene as a function of incident pump fluences. (B) Temperature-dependent chemical potential in graphene following photoexcitation. (C) Distribution of hot electrons for a few typical carrier temperatures $T$. (D) Distribution of hot holes for a few typical carrier temperatures $T$. (E) Estimation of the hot carrier percentage over the energy barrier for HET ($\eta_{electron}$), HHT ($\eta_{hole}$) and their difference ($\eta_{electron} - \eta_{hole}$) at different carrier temperatures.

According to the early study (*3*), the photo-induced THz response of weakly doped graphene ($(-\frac{\Delta E}{E})_g$) is linear proportional to electron temperature $T_e$ and shows a nonlinear dependence on the incident fluence $F$, as described in Eq. (S7),

$$(-\frac{\Delta E}{E})_g = \alpha T_e = \alpha(T_L^3 + \frac{3\gamma F}{\beta})^{1/3} \quad (S7)$$

where $T_L$ is the lattice temperature, $\beta = \frac{18\zeta(3)k_B^3}{(\pi\hbar v_F)^2}$ and $\zeta(3) = 1.202$, $k_B$ is the Boltzmann constant, $v_F$ is the Fermi velocity of graphene and $\hbar$ is the reduced Planck constant. Here we use $\gamma = 0.0161$ (70 % × 2.3 %) on the basis of 70 % of incident photon energy is transferred to the electronic system following 1.55 eV excitation (*15*) and the absorption of graphene at 1.55 eV is 2.3 % (*16*, *17*). Since the photo-induced THz response of the heterostructure ( $(-\frac{\Delta E}{E})_{g-WS_2}$ ) after ~ 1 ps following 1.55 eV excitation (Fig. 2A, red line) is composed of two contributions, i.e., a negative photoconductivity originating from the intrinsic hot carrier response of graphene and a positive photoconductivity due to HET from p-doped graphene to WS$_2$, we use $[(-\frac{\Delta E}{E})_{min, g-WS_2} - (-\frac{\Delta E}{E})_{max, g-WS_2}]$ to approximate the photo-induced THz response for bare graphene, i.e., $(-\frac{\Delta E}{E})_g$, under the same pump energy and fluence. **Figure S3(A)** shows the absolute value of the approximated photo-induced THz response for bare graphene, i.e., $\left|(-\frac{\Delta E}{E})_g\right|$ (as the left Y axis), at different incident fluences and their corresponding $T_e$ (as the right Y axis) obtained by Eq. (S7). $T_e$ shows a sub-linear dependence on $F$ and reaches at 1500 K~3000 K in our employed fluence range. **Figure S3(B)** plots the temperature-dependent chemical potential ($\mu_T$) in graphene following photoexcitation. The chemical potential at different $T$ can be obtained by considering the conservation of the total particle number in graphene and follows $\mu = \frac{4\ln(2)E_F^2}{k_B T}$ according to the early study (*18*).

**Figure S3(C)** plots the electron distribution at typical $T$ according to Eq. (S8),

$$\int_{\mu_T}^{\infty} v(E)f(E)dE \quad (S8)$$

where $f(E) = \dfrac{1}{e^{(E-E_f)/k_B T_e}+1}$ is the Fermi-Dirac distribution and $v(E) = \dfrac{2E}{\pi(\hbar v_F)^2}$ is the density of states (DOS). In our employed fluence range ($T > 1500$ K), we find interband heating takes place effectively and a fraction of hot electrons can be indeed heated up with very high energetics beyond the interfacial barrier for HET (+0.8 eV). This further supports our proposed HET mechanism: a fraction of thermalized hot electrons have sufficient energy to overcome the interfacial energy barrier for HET. Similarly, we also plot the hole distribution at typical $T$ using Eq. (S9),

$$\int_{-\infty}^{\mu_T} v(E)f(E)dE \quad (S9)$$

$$\int_{-\infty}^{\mu_T} v(E)f(E)dE \quad (S9)$$

As shown in **Figure S3(D)**, the amount of holes above the interfacial barrier for HHT is much less than that of electrons above the interfacial barrier for HET under the same $T$, indicating HHT is much less likely happen than HET in our system due to its larger interfacial barrier (-1.3 eV).

The hot carrier percentage over the energy barrier for HET ($\eta_{electron}$) and HHT ($\eta_{hole}$) can be quantified as Eq. (S10) and Eq. (S11), respectively.

$$\eta_{HET,\max} = \dfrac{\int_{0.8}^{\infty} v(E)f(E)dE}{\int_{\mu_T}^{\infty} v(E)f(E)dE} \times 100\% \quad (S10)$$

$$\eta_{HHT,\max} = \dfrac{\int_{-\infty}^{-1.3} v(E)f(E)dE}{\int_{-\infty}^{\mu_T} v(E)f(E)dE} \times 100\% \quad (S11)$$

**Figure S3(E)** compares $\eta_{electron}$ and $\eta_{hole}$ as a function of $T$. As expected, $\eta_{electron}$ is much lower than $\eta_{hole}$ under the same $T$ due to its much larger interfacial barrier.

## Section S6. $α_{TA}$ and $α_{THz-TDS}$ at different photon excitation energies

$α_{TA}$ and $α_{THz-TDS}$ are obtained by fitting fluence dependent PB intensities averaged over the A-exciton resonance between 1.88 eV and 2.07 eV (obtained by TA measurements, **Figure S4**) and fluence dependent injected electron densities (obtained by THz-TDS measurements at a pump-probe delay of ~100 ps, **Figure S5**) under different excitation energies, respectively.

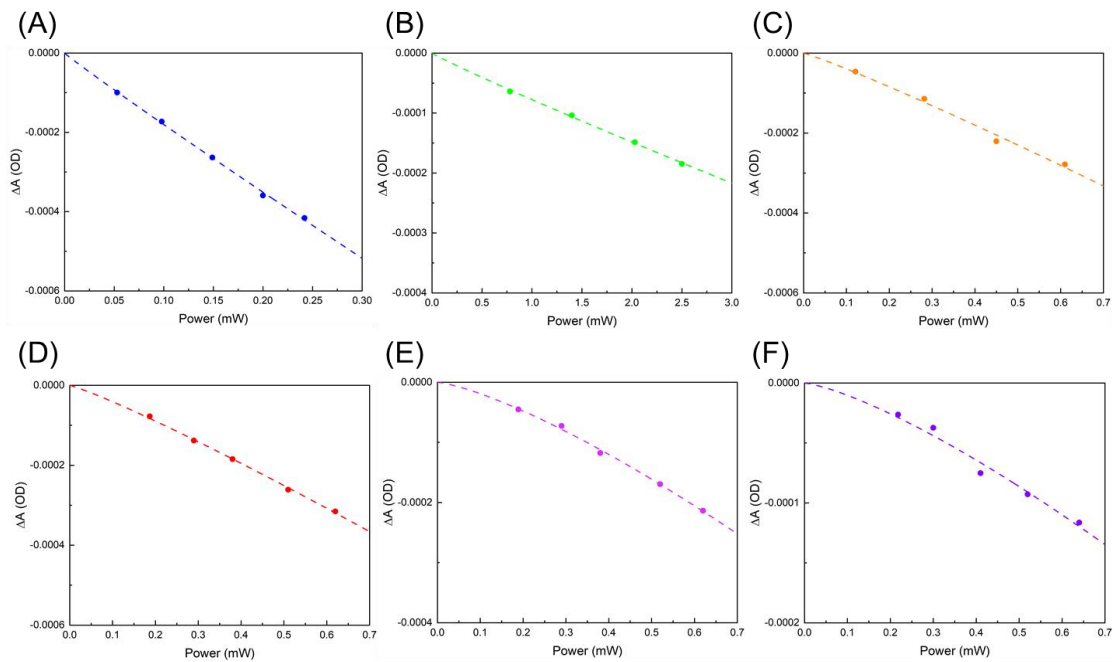

**Figure S4**. Fluence dependent maximum PB intensities (averaged over the A-exciton resonance between 1.88 eV and 2.07 eV) of the heterostructure under (A) 3.10 eV, (B) 2.07 eV, (C) 1.77 eV, (D) 1.55 eV, (E) 1.46 eV and (F) 1.38 eV excitations.

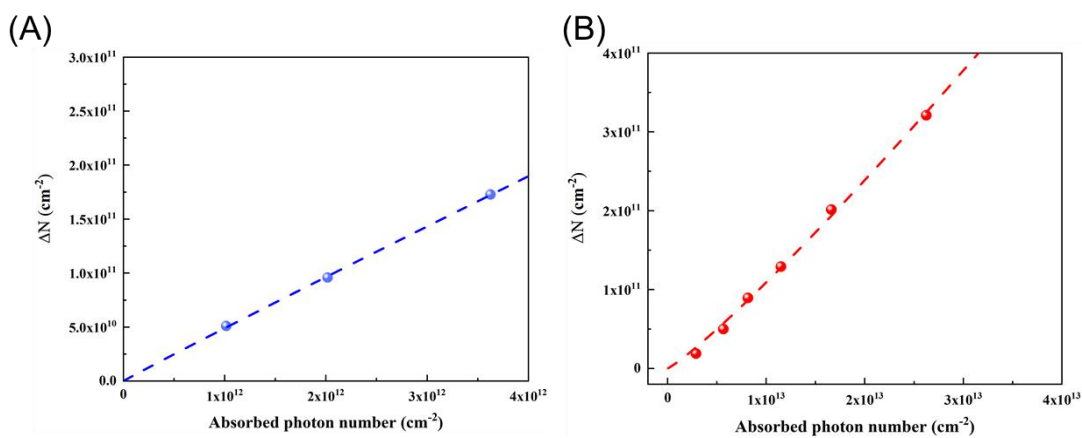

**Figure S5.** Fluence dependent injected electron densities (at a pump-probe delay of ~100 ps) under (A) 3.10 eV, (B) 1.55 eV excitations.

**Section S7. Discussion on ET pathways involving photoexcited hot electrons in graphene in the above-A-exciton excitation regime**

Here we extend our discussions on alternative mechanisms (rather than DHT proposed in the paper) in the above-A-exciton excitation regime, which involves the photoexcited hot electrons in the graphene.

First, we discuss photo-thermionic emission: the mechanism governing the ET for sub-A-exciton excitation as discussed in the paper. We argue that this process will be suppressed by an increase of charging energy in $WS_2$ (*11*) and thus the hot electron injection barrier due to the presence of directly-photogenerated electrons in $WS_2$. This argument is indirectly supported by observing a linear (or slightly sublinear) trend in the fluence-dependent signals with above-A-exciton excitation. This is in contrast to the superlinear dependence expected for photo-thermionic emission.

Second, we have also considered other ET pathways, e.g., via hole transfer from the $WS_2$ valence band to the photoexcited hot electrons in graphene. The contribution of such ET scenario will be small, due to the much lower density of hot electrons in comparison to the "cold" ones in the valence band of graphene ($10^{11}$-$10^{12}$ cm$^{-2}$ *vs* $10^{14}$ cm$^{-2}$).

On top of these discussions, as mentioned already in the paper, the overall contribution of hot electron in graphene will be very small, due to the weak absorption in graphene, in comparison to the strong absorption in the exciton states in $WS_2$.

**Section S8. Photon energy-dependent maximum positive photoconductivity normalized to the incident power**

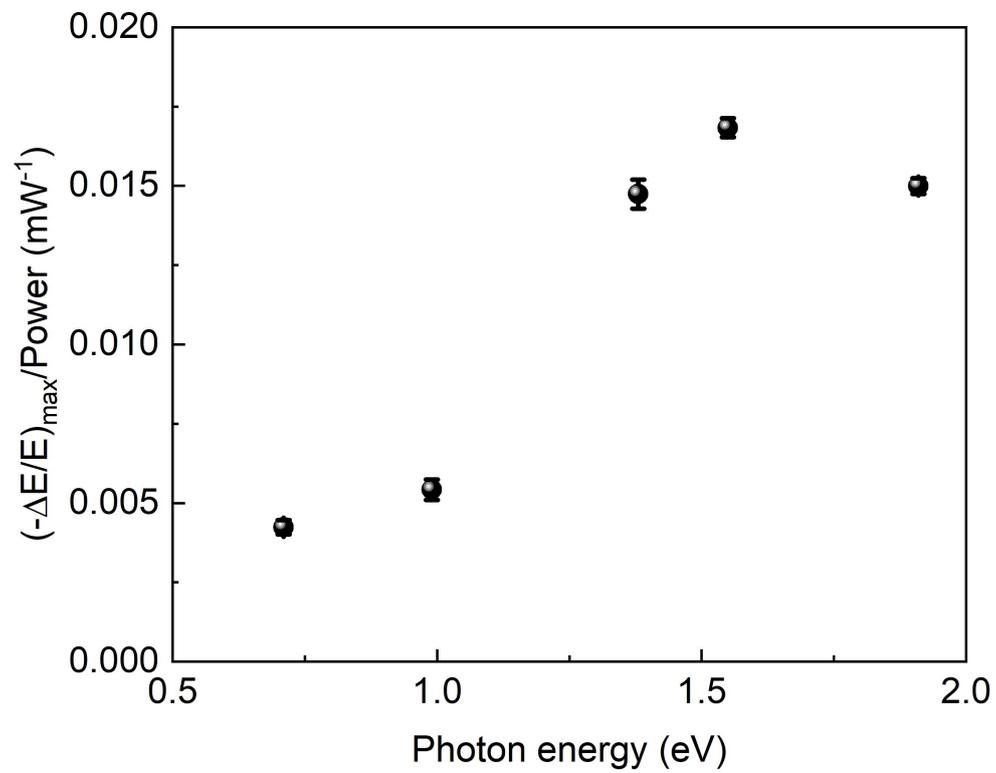

**Figure S6.** Photon energy-dependent maximum positive photoconductivity normalized to the incident power

# Section S9. Normalized photon excitation energy dependent PB dynamics in the g-WS$_2$ heterostructure

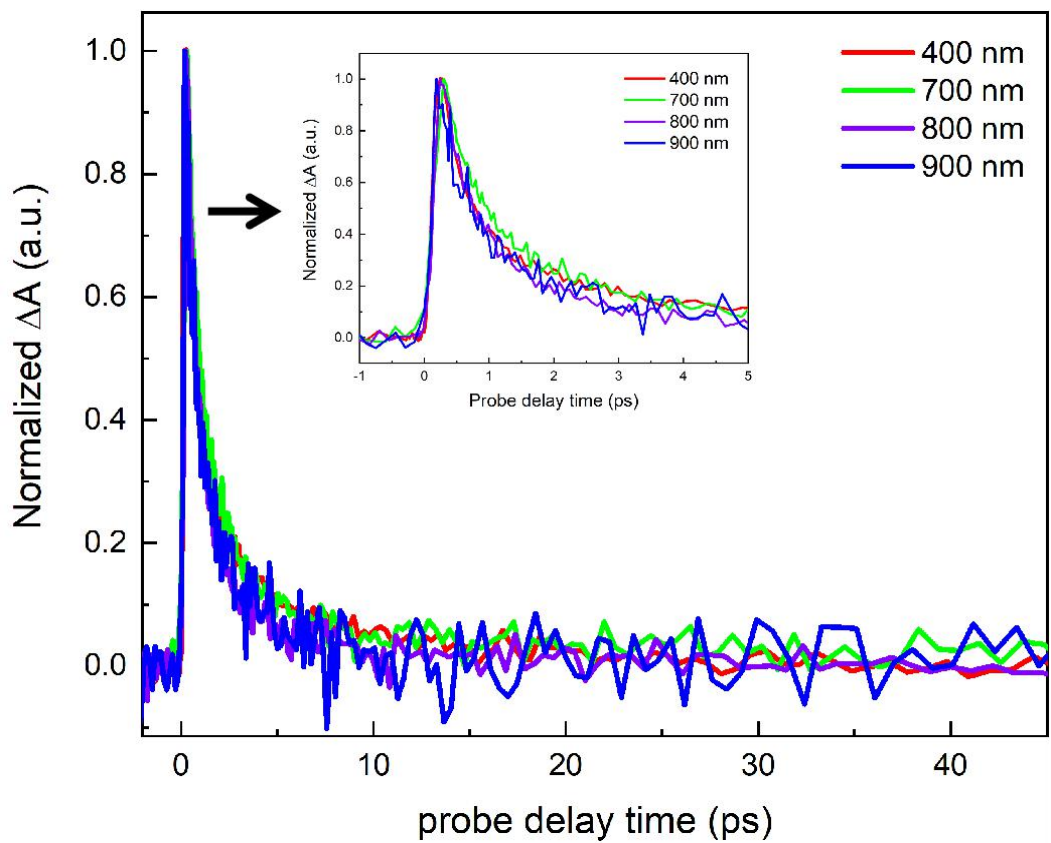

**Figure S7.** Normalized photon excitation energy dependent PB dynamics (at A-exciton resonance of WS$_2$) in the g-WS$_2$ heterostructure. Inset shows the dynamics for the initial time period of 5 ps.

**Section S10. TA spectrum for the g-WS$_2$ heterostructure averaged at different pump probe delay times upon 1.55 eV excitation**

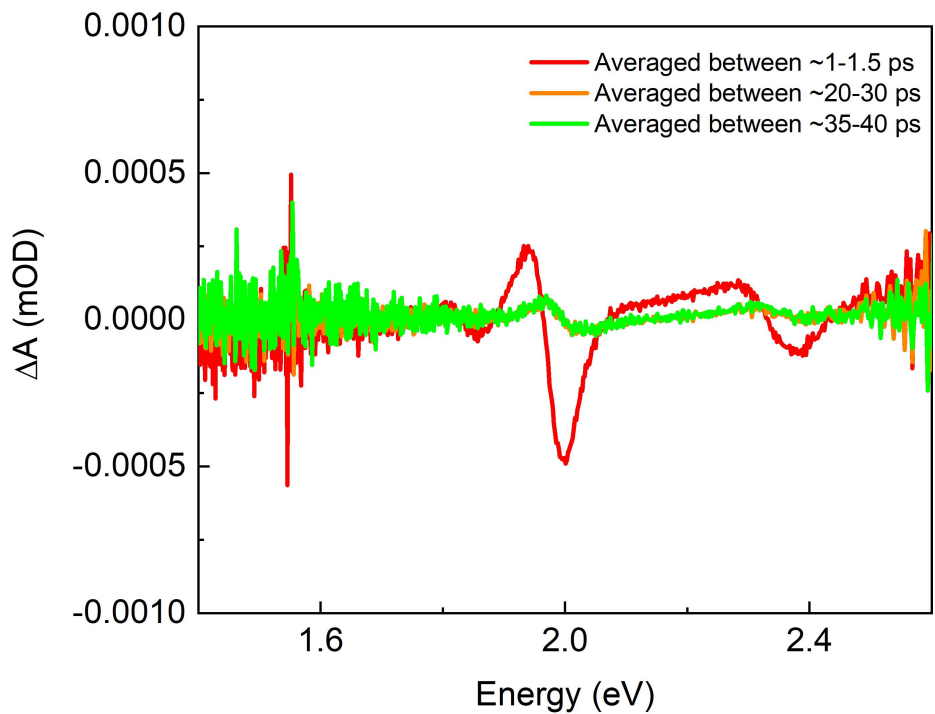

**Figure S8.** TA spectrum for the g-WS$_2$ heterostructure averaged at different time periods upon 1.55 eV excitation with a pump fluence of 620 µW.